# Phase Noise Model for Continuous-Variable Quantum Key Distribution Using a Local Local Oscillator

Yun Shao,[1] Heng Wang,[1] Yaodi Pi,[1] Wei Huang,[1] Yang Li,[1] Jinlu Liu,[1] Jie Yang,[1] Yichen Zhang,[2] and Bingjie Xu[1,*]

[1]*Science and Technology on Communication Security Laboratory, Institute of Southwestern Communication, Chengdu 610041, China*

[2]*State Key Laboratory of Information Photonics and Optical Communications, Beijing University of Posts and Telecommunications, Beijing 100876, China*

*xbjpku@pku.edu.cn

The value of residual phase noise, after phase compensation, is one of the key limitations of performance improvement for continuous-variable quantum key distribution using a local local oscillator (LLO CV-QKD) system, since it is the major excess noise. However, due to the non-ideality of the realistic devices implemented in practice, for example, imperfect lasers, detectors and unbalanced interferometers, the value of residual phase noise in current system is still relatively large. Here, we develop a phase noise model to improve the phase noise tolerance of the LLO CV-QKD schemes. In our model, part of the phase-reference measurement noise associated with detection efficiency and electronic noise of Bob's detector as well as a real-time monitored phase-reference intensity at Bob's side is considered trusted because it can be locally calibrated by Bob. We show that using our phase noise model can significantly improve the secure key rate and transmission distance of the LLO CV-QKD system. We further conduct an experiment to substantiate the superiority of the phase noise model. Based on experimental data of a LLO CV-QKD system in the 25 km optical fiber channel, we demonstrate that the secure key rate under our phase noise model is approximately 40% higher than that under the conventional phase noise model.

# I. INTRODUCTION

Continuous-variable quantum key distribution (CV-QKD) has attracted considerable attention over the past decades, in which the secret keys are encoded in quadratures of coherent or squeezed states, and decoded by homodyne or heterodyne detections [1-3]. Benefiting from the utilizing of coherent detection, CV-QKD is allowed to be fully implemented with the state-of-the-art optical telecom components, which are well compatible with the existing telecom industry [4, 5]. Several seminal CV-QKD protocols have been proposed. In particular, Gaussian-modulated coherent-state (GMCS) CV-QKD protocol [6, 7], as the most widely developed CV-QKD protocol, has been investigated in depth [2, 3] and experimentally demonstrated with a long distance CV-QKD over 202.81 km optical fiber channel [8]. Although some fantastic progress has been made in CV-QKD, there still remains an ongoing challenge due to coherent detection which usually requires the sender to generate the local oscillator (LO) and the quantum signal from the same laser and co-transmit them through the insecure quantum channel [9]. This implementation leaves some security loopholes, which can be exploited by the eavesdropper to mount attacks, such as the wavelength attack [10-12], LO fluctuation attack [13], LO calibration attack [14], LO polarization attack [15], and jitter in clock synchronization [16]. Meanwhile, some corresponding countermeasures have been proposed to resist the attacks. For example, a wavelength filter can be applied to counter the wavelength attack, and a real-time shot noise measurement can be performed to prevent all LO intensity attack [17, 18]. In addition to the LO attacks, the detector saturation attack [19] and homodyne detector blinding attack [20] were also demonstrated in recent works. Moreover, the shot-noise-limited homodyne detection of quantum states should be performed with quite a strong LO, which may result in the leakage of photons from LO to signal and dramatically reduce the efficiency of CV-QKD [21].

Recently, a novel Local LO scheme for CV-QKD (LLO CV-QKD) has been proposed [22, 23] and successfully demonstrated, in which the transmission of LO through the insecure quantum channel is completely avoided. Hence, it is no surprise that any security loopholes associated with manipulating the LO have been effectively eliminated, and the photons leakage from LO to signal has also been blocked. In LLO CV-QKD, the quantum signal is generated on Alice's side, while the LO is generated using an independent laser on Bob's side. In order to establish a reliable phase reference between the two independent free running lasers, a low intensity classical phase-reference needs to be transmitted together with the signal pulse. Several phase-reference schemes have been introduced to promote the efficiency of LLO CV-QKD, such as LLO sequential scheme [22-24], LLO delay-line scheme [25], LLO pilot-multiplexed scheme [26, 27] and LLO pilot-tone-assisted scheme [28]. So far, in the asymptotic limit, LLO CV-QKD experiments have been achieved with a secure key rate up to 7.04 Mbits/s over 25 km optical fiber [28] and 26.9 Mbits/s over 15 km optical fiber [29].

In most practical LLO CV-QKD schemes, since the signal pulse and the LO are generated from two independent lasers, the excess noise is relatively large, which greatly limits the secure key rate and transmission distance of the system. As a crucial component in determining the excess noise, the phase noise is conventionally considered as untrusted noise. Jouguet et al. [30] introduced a realistic model to derive a better secure key rate, in which the phase noise was expected to be calibrated locally and considered as trusted. More recently, a refined noise model has been proposed to improve the phase noise tolerance of the LLO CV-QKD scheme [31], in which the phase noise associated with coherent detector is regarded as trusted noise while the phase noise caused by the relative phase drift between the signal pulse and the phase-reference is regarded as untrusted noise. Ren et al. [32] treats the phase noise associated with measurement of the phase-reference as trusted noise, which dramatically improves the performance of the LLO CV-QKD system. However, the trusted part of phase noise is overestimated, resulting in an overestimation of secure key rate and transmission distance. In this paper, the authors also presented an interesting attack by manipulating the phase-reference intensity. As a countermeasure, they suggest that monitoring the phase-reference intensity in real-time is effective.

In this work, we develop a phase noise model for LLO CV-QKD protocols. The assumption in our phase noise model is that part of the phase noise related to detection efficiency and electronic noise of Bob's detector as well as a real-time monitored phase-reference intensity at Bob's side is trusted noise. It is reasonable to adopt this model because the above parameters can be locally calibrated by Bob. Using this phase noise model, we present the simulation results of the secure key rate of the LLO pilot-multiplexed CV-QKD system with some typical parameters. We show that the LLO CV-QKD performance is significantly improved compared with that under the conventional phase noise model. Our phase noise model can be applicable to all current LLO CV-QKD protocols with phase-reference since the phase noise is inevitably introduced during the QKD process. We further conduct a LLO pilot-tone-assisted CV-QKD experiment to quantitatively confirm the advantage of the phase noise model.

This paper is organized as follows. In Section II, we review the LLO CV-QKD protocol based on Gaussian modulated coherent states and heterodyne detection, and we present the expressions for the residual phase noise of the LLO CV-QKD system after phase compensation. In Section III, we develop a trusted phase noise model and compare the secure key rate of LLO pilot-multiplexed CV-QKD system under different phase noise models. We also consider and discuss extensively the security of a practical LLO CV-QKD system under phase-reference intensity attack. In Section IV, we conduct an experiment to demonstrate the advantage of the trusted phase noise model. Finally, we conclude the results of our work in Section V.

## II. PHASE NOISE IN LLO CV-QKD

In this section, we first review the LLO CV-QKD protocol based on Gaussian-modulated coherent states and heterodyne detection, and then we present the expressions for the residual phase noise after phase compensation.

### A. LLO CV-QKD Scheme

Coherent detection of the quantum signal requires a high-power LO and a stable relative phase between the quantum signal and the LO at reception. In the transmitted LO CV-QKD scheme, both the weak quantum signal and the strong LO are generated from the same laser on Alice's side and propagated on the public channel to guarantee a stable relative phase.

However, for a typical GMCS LLO CV-QKD protocol [22-24], as shown in Fig. 1, Bob adopts a second independent laser to generate LO locally for coherent detections. At the sender, Alice prepares two independent Gaussian random variables $X_A$ and $P_A$ which obey the same zero-centered Gaussian distribution $\mathcal{N}(0, V_A)$, where $V_A$ is the modulation variance. Then Alice sends the Gaussian-modulated coherent state $|\alpha_S\rangle = |X_A + iP_A\rangle$, referred to as quantum signal, to Bob through the insecure quantum channel. In order to create a reliable relative phase between the signal pulse and the LO pulse, she also sends a classical coherent state $|\alpha_R\rangle = |X_A^R + iP_A^R\rangle$, referred to as phase-reference, to Bob along with the signal pulse. Both the signal pulse and the phase-reference are generated from the same laser. After repeating this process many times, an interleaved signal pulse and phase-reference pulse are simultaneously transmitted from Alice to Bob. In fact, the intensity of the phase-reference is constant and relatively larger than the signal pulse while much weaker than the LO [23]. This arrangement allows one to reduce the interference between the signal pulse and the phase-reference in the multiplexed CV-QKD system. At the receiver, Bob locally generates a high-power LO and splits it into two beams for the coherent detections of signal pulse and phase-reference, respectively. He performs a heterodyne detection to simultaneously measure the quadratures of signal pulse and obtains the outcomes $X_B$ and $P_B$, and a heterodyne detection to simultaneously measure the quadratures of phase-reference and obtains the outcomes $X_B^R$ and $P_B^R$. According to the measurement outcomes of the phase-reference, the phase rotation between the two free-running lasers is estimated as $\hat{\theta}_R = \tan^{-1}(P_B^R/X_B^R)$ [23].

Thus, during the reverse reconciliation, Alice gets the phase-rotation estimation value of the phase-reference from Bob and estimates Bob's measurement results of the quantum signal $X_B'$ and $P_B'$ by performing a phase rotation $-\hat{\theta}_R$ on her quantum signal to complete the phase compensation[25, 32]

$$\begin{pmatrix} X_B' \\ P_B' \end{pmatrix} = \sqrt{\frac{T\eta}{2}} \begin{pmatrix} \cos\hat{\theta}_R & \sin\hat{\theta}_R \\ -\sin\hat{\theta}_R & \cos\hat{\theta}_R \end{pmatrix} \begin{pmatrix} X_A \\ P_A \end{pmatrix}. \qquad (1)$$

Here, $\eta$ is the detector efficiency, T is the quantum channel transmittance, given by $T = 10^{-\alpha d/10}$, where d is the transmission distance, and $\alpha$ is the loss coefficient for fiber channel. After the phase compensation procedure, Alice and Bob can share a partially correlated Gaussian random variable. The remaining processes are performed to complete the LLO CV-QKD protocol, including parameter estimation, error correction, and privacy amplification.

## B. Phase noise

It is worth noting that, in practice, the phase rotation of the phase-reference is not strictly equal to that of the signal pulse, there exists an error, i.e., residual phase noise after the phase compensation procedure. To illustrate the source of phase noise, we analyze its composition. We define $\theta_S$ as the real phase rotation value between the signal pulse and the LO while $\hat{\theta}_S$ as its estimated value. We define $\theta_R$ as the real phase rotation value between the phase-reference and the LO while $\hat{\theta}_R$ as its estimated value. In the GMCS protocol with modulation variance $V_A$, the phase noise can be expressed as [25]

$$\xi_{\text{phase}} = 2V_A\left(1 - e^{-\frac{V_{est}}{2}}\right). \qquad (2)$$

Here, $V_{est}$ is the variance of the phase noise, which comes from the difference between the actual phase rotation value of the signal pulse and its estimation value,

$$V_{est} = var(\theta_S - \hat{\theta}_S), \qquad (3)$$

According to the steps of the relative phase-rotation estimation process, the phase noise variance mainly consists of three parts [25, 27] (see Fig. 1),

$$V_{est} = V_{drift} + V_{channel} + V_{error}, \qquad (4)$$

The first term is the variance of the relative phase drift between the signal pulse and the phase-reference at emission [25],

$$V_{drift} = 2\pi(\Delta\nu_A + \Delta\nu_B)|t_R - t_S|, \qquad (5)$$

where $\Delta\nu_A$ and $\Delta\nu_B$ correspond to the linewidths of the two free-running lasers, and $t_S$ and $t_R$ correspond to the emission time of the signal pulse and the phase-reference.

The second term is mainly caused by the drift of phase-accumulation between the signal pulse and the phase-reference during their propagation [25],

$$V_{channel} = var(\theta_S^{ch} - \theta_R^{ch}). \qquad (6)$$

where $\theta_S^{ch}$ and $\theta_R^{ch}$ respectively represent the phases accumulated by the signal pulse and the phase-reference.

Considering the basic shot noise and the experimental noise in heterodyne detection, the third term represents the variance caused by measurement of the phase-reference,

which is characterized by [23, 25]

$$V_{error} = var(\theta_R - \hat{\theta}_R) = \frac{\chi + 1}{E_R^2}. \tag{7}$$

with $E_R$ being the phase-reference amplitude on Bob's side, and $\chi$ being the total noise imposed on the phase-reference, given by [25, 27]

$$\chi = \frac{1-T}{T} + \varepsilon_0 + \frac{2 - \eta + 2v_{el}}{T\eta}. \tag{8}$$

In Eq. (8), the first term stands for the loss-induced vacuum noise, the second term is the excess noise of the phase-reference from the channel, with $\varepsilon_0 = 0.002$ (shot noise unit, SNU) [9], and the third term stands for the noise added by the imperfection of phase-reference heterodyne detection. Thus, in low $V_{est}$, the phase noise $\xi_{phase}$ can be expressed as [25, 27],

$$\xi_{phase} = V_A V_{est} = V_A(V_{drift} + V_{channel} + V_{error}) = \xi_{drift} + \xi_{channel} + \xi_{error}. \tag{9}$$

### III. TRUSTED PHASE NOISE MODEL FOR LLO CV-QKD

#### A. Trusted phase noise model

The excess noise $\xi_{tot}$ is a key parameter to evaluate the performance of CV-QKD system, which can be calibrated in the parameter estimation process. Besides the phase noise $\xi_{phase}$, the other noise $\xi_{rest}$ will also undermine the system performance and contributes to the excess noise [25-27]. Thus, the excess noise in the LLO CV-QKD system can be expressed as

$$\xi_{tot} = \xi_{phase} + \xi_{rest}. \tag{10}$$

In a practical CV-QKD system for heterodyne detection, the insecure quantum channel is characterized by the channel transmittance T and the excess noise $\xi_{tot}$. The detector is characterized by the efficiency $\eta$ and the electronic noise $v_{el}$. The total added noise imposed on the signal pulse referred to the channel input between Alice and Bob can be modeled as $\chi_{tot} = \chi_{line} + \chi_{het}/T$, where $\chi_{line}$ represents the total channel added noise referred to the channel input which can be defined as $\chi_{line} = 1/T - 1 + \xi_{tot}$ [33]. $\chi_{het}$ represents the total detection added noise referred to Bob's input, which can be defined as $\chi_{het} = [1 + (1-\eta) + 2v_{el}]/\eta$ [33]. Recently, a robust trusted detector noise model has been adopted in CV-QKD [31-35], in which the parameters $\eta$ and $v_{el}$ can be well calibrated locally at the receiver's side. Thus, the detection added noise $\chi_{het}$ can be regarded as trusted noise, which is consistent with the assumption that Bob's apparatus is inaccessible to Eve.

Based on the trusted detector noise model and the previous refined noise model [29-35], we next introduce a trusted phase noise model which could be applied to current LLO CV-QKD schemes with phase-reference. Similar to Ref. [31, 32], in our model, the relative phase drift noise $\xi_{drift}$ and the relative phase accumulated noise $\xi_{channel}$ are also regarded as untrusted noise. Nevertheless, part of the phase-reference measurement noise $\xi_{error}$ can be well calibrated locally on Bob's side is thus can be regarded as trusted noise.

According to Eq. (8), the total noise imposed on the phase-reference can be divided into two parts:

$$\chi = \chi^u + \frac{\chi^T}{T}. \tag{11}$$

with $\chi^u = 1/T - 1 + \varepsilon_0$ and $\chi^T = (2 - \eta + 2\nu_{el})/\eta$. Therefore, in the LLO CV-QKD scheme, the measurement noise of the phase-reference can be decomposed into two categories, namely [according to Eqs. (7)–(9) and (11)],

$$\xi_{error} = V_A V_{error} = \xi^u_{error} + \frac{\xi^T_{error}}{T}. \tag{12}$$

where

$$\xi^u_{error} = V_A \left(\frac{\chi^u + 1}{E_R^2}\right) = V_A \left(\frac{1 + T\varepsilon_0}{T E_R^2}\right), \tag{13}$$

$$\xi^T_{error} = V_A \left(\frac{\chi^T}{E_R^2}\right) = V_A \left(\frac{2 - \eta + 2\nu_{el}}{\eta E_R^2}\right). \tag{14}$$

Here, $\xi^T_{error}$ represents part of the phase-reference measurement noise referred to Bob's input, which is determined by the detector efficiency $\eta$ and the detector electronic noise $\nu_{el}$, as well as the phase-reference intensity $E_R^2$ on Bob's side. It can be well calibrated by Bob at the receiver side as Bob's apparatus is inaccessible to Eve. Given that the phase-reference propagates through the insecure quantum channel, this may leave security loopholes for the eavesdropper. According to Eq. (14), although the eavesdropper can mount attacks by manipulating the intensity of the phase-reference, a real-time calibration of $\xi_{error}^T$ by monitoring the instantaneous intensity of the phase-reference on Bob's side can prevent the attack and guarantee the security of the LLO CV-QKD system. Therefore, the real-time calibrated $\xi_{error}^T$ can be regarded as trusted noise. In a practical system, Bob can introduce a beam splitter, or a quick random switch to switch the phase-reference pulse to a high sensitivity power meter to monitor the intensity of the phase-reference. In practice, in order to avoid overestimating the trusted part of the phase noise, we can use the upper bound value of the phase-reference intensity fluctuation to calibrate $\xi_{error}^T$, so as to obtain a lower bound value of the secure key rate. Meanwhile, it is conceived that the phase noise $\xi_{error}^u$ including the parameters T and $\varepsilon_0$ which can be controlled by Eve during the QKD run is thus regarded as untrusted noise. According to the above description, one can find that in the trusted phase noise model, the total channel added noise referred to the channel input based on heterodyne detection can be written as

$$\chi_{line}^T = \frac{1}{T} - 1 + \xi_{rest} + \xi_{drift} + \xi_{channel} + \xi_{error}^u$$
$$= \frac{1}{T} - 1 + \xi_{rest} + \xi_{phase} - \frac{\xi_{error}^T}{T}$$
$$= \frac{1}{T} - 1 + \xi_{tot}^T, \quad (15)$$

where $\xi_{tot}^T = \xi_{tot} - \xi_{error}^T/T$ is the real excess noise under the trusted phase noise model. Moreover, the total detection added noise referred to Bob's input can be written as

$$\chi_{het}^T = \frac{2 - \eta - 2v_{el}}{\eta} + \xi_{error}^T, \quad (16)$$

Therefore, the total added noise referred to the channel input between Alice and Bob can be given by

$$\chi_{tot}^T = \chi_{line}^T + \frac{\chi_{het}^T}{T}. \quad (17)$$

As a comparison, the corresponding terms in the conventional phase noise model (all sources of the phase noise are regarded as untrusted noise) can be written as

$$\chi_{line} = \frac{1}{T} - 1 + \xi_{rest} + \xi_{phase},$$
$$\chi_{het} = \frac{2 - \eta + 2v_{el}}{\eta},$$

$$\chi_{\text{tot}} = \chi_{\text{line}} + \frac{\chi_{\text{het}}}{T}. \tag{18}$$

**B. Calculations and numerical simulations**

For simplicity of description, in this section, we consider the LLO pilot-multiplexed scheme based on heterodyne detection [27]. In this case, the signal pulse and the phase-reference are generated from the same optical wave front, i.e., $t_R = t_S$, according to Eq. (5), $V_{\text{drift}}$ can be ignored. Meanwhile, the relative phase accumulated variance is a small amount and can be regard as $V_{channel} \approx 0$ [25]. Therefore, the phase noise in Eq. (9) can be simplified as the phase-reference measurement noise,

$$\xi_{\text{phase}} \approx V_A V_{error} = \xi_{error} = V_A \left( \frac{\chi + 1}{E_R^2} \right). \tag{19}$$

According to Eqs. (15)−(17), here the added noises in the trusted phase noise model can be written as

$$\chi_{line}^T = \frac{1}{T} - 1 + \xi_{rest} + \xi_{error} - \frac{\xi_{error}^T}{T},$$

$$\chi_{het}^T = \frac{2 - \eta + 2v_{el}}{\eta} + \xi_{error}^T = \frac{2 - \eta + 2v_{el}}{\eta} \left( 1 + \frac{V_A}{E_R^2} \right),$$

$$\chi_{tot}^T = \chi_{line}^T + \frac{\chi_{het}^T}{T}. \tag{20}$$

The components of the excess noise term $\xi_{rest}$ are detailed in Appendix A. The secure key rate of the LLO CV-QKD system under the collective attack can be calculated based on Appendix B. From Eqs. (18) and (20), one can find that the mutual information between Alice and Bob in the trusted phase noise model is equal to that in the conventional phase noise model, i.e.,

$$I_{AB}^T = \log_2 \frac{V + \chi_{tot}^T}{1 + \chi_{tot}^T} = \log_2 \frac{V + \chi_{tot}}{1 + \chi_{tot}} = I_{AB}. \tag{21}$$

Therefore, the secure key rate discrepancy of the LLO CV-QKD system between the trusted phase noise model and the conventional phase noise model is determined by the Holevo bound value.

In the conventional phase noise model, different from that in the trusted phase noise model where part of the phase noise is considered trusted, all sources of the phase noise is supposed to be untrusted, which makes the phase noise tolerance of the system very low. Figure 2 depicts the simulation results of the secure key rate versus the transmission distance under the conventional phase noise model (blue dash-dotted line) and the trusted phase noise model (solid green line). One can observe that the LLO CV-QKD performance under the trusted phase noise model is significantly improved compared with that under the conventional phase noise model. With some typical parameters conditions [26, 31, 32], the maximum transmission distance is anticipated to promote over 65%. For the secure key rate, taking the simulation results of 25 km transmission distance as an example, we find that it is increased by more than 60%. This underlines the fact that the conventional phase noise model is too pessimistic. For comparison, we further show the result under the phase noise model of Ref. [32] (dashed red line). In that work, they treated the total phase-reference measurement noise $\xi_{\text{error}} = V_A(\chi + 1)/E_R^2$ as trusted noise, which overestimates the trusted part of the phase noise, and rendering an optimistic overestimation of the secure key rate and the transmission distance.

It is interesting to note that the phase noise can be reduced by increasing the intensity of the phase-reference [according to Eq. (19)]. However, the reduction of the phase noise comes at the expense of increasing interference between the signal pulse and the phase-reference in the multiplexed LLO CV-QKD system. Therefore, the choice of the phase-reference intensity is the result of a trade-off, meanwhile, the introduction of the phase noise is inevitably.

To further substantiate our claim that the trusted phase noise model can resist the previous attacks in Ref. [32] and improve the secure key rate and the transmission distance, we consider the security of practical LLO CV-QKD system under the trusted phase noise model of monitoring the phase-reference intensity in real-time, and calculate the secure key rate under the attack. As described by Ren et al., in a practical attack scheme, Eve can lower the trusted part of the phase noise by increasing the intensity of the phase-reference to compensate for her increased attack noise on the signal pulse so as to keep the total noise of Bob's measurement constant. In this case, Eve can transmit the phase-reference pulse using an ultralow-loss optical fiber with $\alpha_{\text{low}} = 0.14$ dB/km and transmit the signal pulse with another standard optical fiber with $\alpha_{\text{std}} = 0.2$ dB/km, which is equivalent to increasing the phase-reference intensity received on Bob's side.

Therefore, the added noises of the LLO CV-QKD system in the trusted phase noise model under the phase-reference intensity attack can be written as

$$\chi_{line}^{T} = \frac{1}{T} - 1 + \xi_{rest} + \xi_{error} - \frac{\xi_{error}^{T}}{T} + \frac{\xi_{attack}}{T},$$

$$\chi_{het}^{T} = \frac{2-\eta+2\nu_{el}}{\eta} + \xi_{error}^{T} - \xi_{attack} = \frac{2-\eta+2\nu_{el}}{\eta}\left(1 + \frac{V_A}{E_{R\_attack}^2}\right),$$

$$\chi_{tot}^{T} = \chi_{line}^{T} + \frac{\chi_{het}^{T}}{T}. \tag{22}$$

where $\xi_{attack}/T$ is the excess noise introduced by Eve's increased attack on the signal pulse, and $E_{R\_attack}^2$ is the phase-reference intensity on Bob's side under attack. $\xi_{attack}$ is equal to the reduction of the trusted part of the phase noise and given by

$$\xi_{attack} = V_A\left(\frac{\chi^T}{E_R^2}\right) - V_A\left(\frac{\chi^T}{E_{R\_attack}^2}\right) = \xi_{error}^T\left(1 - \frac{1}{10^{(\alpha_{std}-\alpha_{low})*L/10}}\right) \tag{23}$$

where $E_{R\_attack}^2$ and $E_R^2$ respectively represents the phase-reference intensity on Bob's side with and without attack.

In Fig. 2, the black dotted line represents the secure key rate of the trusted phase noise model under the attack. One can find that the performance of the LLO CV-QKD system under the trusted phase noise model is much better in compassion to that under the conventional phase noise model even the LLO CV-QKD system is subjected to the phase-reference intensity attack. In our model, the phase-reference intensity attack can be resisted as the intensity of the phase-reference is monitored in real-time, and Bob can calibrate the trusted part of the phase noise in real-time to obtain a secure key rate. Hence, the trusted phase noise model in combination with monitoring the phase-reference intensity in real-time can prevent the attack and guarantee the security of the LLO CV-QKD system, as well as improve the secure key rate performance. The security analysis above is valid for the phase-reference intensity attack. It is worth noting that the above analysis only shows the result of the trusted phase noise model under a powerful phase-reference intensity attack. We cannot rule out that in some extreme situations Eve may has a way to lower the trusted part of the phase noise severely. In this case, the fluctuation of the phase-reference intensity is considerable, and the communication parties can detect the attack and stop the QKD.

In Fig. 3, we present the simulation results of the excess noise and the trusted part of the phase noise referred to the channel input with respect to the transmission distance. The results show that both of them are proportional to the transmission distance, which means that under the conventional phase noise model, the realization of high key rate and long-distance LLO CV-QKD is greatly limited. More importantly, as the performance of the system is sensitive to the excess noise, a slight increase of the excess noise may prominently degrade the secure key rate. To achieve a high key rate and long-distance LLO CV-QKD, there are rigorous technological challenges and expensive devices cost still existed in suppressing the excess noise. Due to the excess noise in the LLO CV-QKD system with phase-reference is mainly caused by the phase noise, considering the phase noise related to coherent detector and phase-reference intensity that can be calibrated at the receiver's side as trusted noise can significantly improve the phase noise tolerance of the system. To this end, we can conclude that the use of the trusted phase noise model not only has important implications in the achieving of the high key rate and long-distance LLO CV-QKD, but also can help reduce the cost of implementation.

## IV. EXPERIMENTAL DEMONSTRATION

Based on the above analysis, we can find that the trusted phase noise model is applicable to all current LLO CV-QKD schemes with phase-reference because the phase-reference measurement noise is inevitably introduced during the phase-rotation estimation process. In the following, we will conduct a LLO pilot-tone-assisted experiment to present the secure key rate results of the system under the trusted phase noise model and the conventional phase noise model.

Experimentally, we use the Gaussian-modulated LLO pilot-tone-assisted CV-QKD scheme, which is a frequency-polarization multiplexing system [see Ref. [28] for more details]. As shown in Fig. 4, Alice splits the continuous optical carrier into two beams using an unbalanced beam splitter. The weaker optical carrier is modulated by an amplitude modulator to be signal pulse with pulse duration of 2 ns and a repetition of 100 MHz, and then is modulated with Gaussian distribution. The stronger pilot-tone-assisted, i.e., the phase-reference, is modulated by a Mach-Zehnder modulator to form carrier suppression double sideband optical carrier whose frequency is different from that of the signal pulse. Next, a polarization beam combiner is used to recombine the signal pulse and the phase-reference. At the receiver, a polarization controller is used to eliminate the deterioration of the polarization. Then, the frequency-polarization multiplexing optical carrier is separated into signal pulse and phase-reference through a polarization beam splitter. Meanwhile, the LO is locally generated and split into two beams. Bob employs two identical balanced detectors (Thorlabs, PDB 480C) with a band width of 1.6 GHz to perform heterodyne detection on signal pulse and phase-reference respectively. The measurement results of the phase-reference allows Bob to recover the phase rotation between the two free-running lasers.

The experimental parameters can be found in Table I, the transmission distance $L = 25$ km, the attenuation coefficient $\alpha = 0.2$ dB/km, the repetition frequency $f_{rep} = 100$MHz, the detector efficiency $\eta = 0.56$, the reconciliation efficiency $\beta = 0.95$, the electronic noise $v_{el} = 0.042$, and the modulation variance $V_A = 3.073$. The intensity of the phase-reference on Bob's side can be measured by a power meter in advance of the QKD, which is around $E_R^2 = \langle N_R \rangle = 1000$, $\langle N_R \rangle$ is the average photons number of the phase-reference on Bob's side. The mean of the measurement excess noise under the conventional phase noise model is around $\xi_{tot} = 0.056$, which based on the block size of $5 \times 10^6$. As a comparison, the excess noise under the trusted phase noise model is around $\xi_{tot}^T = 0.03$. All noise variances here are expressed in shot noise units (SNU).

As the signal pulse and the phase-reference are generated from the same optical wave front in the LLO pilot-tone-assisted CV-QKD system, the phase noise is also determined by the phase-reference measurement noise. Combining Eqs. (13) and (15)−(18) as well as the calculations in Appendix B, the secure key rate of the practical CV-QKD system could be $Key = f_{rep} \cdot K$. With the present experimental parameters, one can calculate the secure key rate of the system under the collective attack. The secure key rates of experimental results and numerical simulations are shown in Fig. 5. The results show that the experimental secure key rates at 25 km under the conventional phase noise model and the trusted phase noise model are $Key = 4.556$Mbps and $Key^T = 6.358$Mbps, respectively. It indicates that the secure key rate can be increased by approximately 40% when using the trusted phase noise model compared with that using the conventional phase noise model. The numerical simulated results show the significant improvement of the LLO CV-QKD system under the trusted phase noise model both in secure key rate and transmission distance in comparison to that under the conventional phase noise model.

## V. CONCLUSION

In conclusion, we have developed a trusted phase noise model which can be applicable to current LLO CV-QKD protocols with phase reference. Using this model, we present simulation results to show that the secure key rate and the transmission distance of the LLO CV-QKD system is significantly improved. We also show the trusted phase noise model with monitoring the phase-reference intensity in real-time can prevent the attacks and guarantee the security of the system. We have demonstrated an experimental scheme which reveals the superiority of the trusted phase noise model. We show that the secure key rate of the practical LLO CV-QKD system at the transmission distance of 25 km optical fiber channel under the trusted phase noise model is improved by approximately 40% compared with that under the conventional phase noise model. The present study will facilitate to refine the phase noise model to improve the phase noise tolerance of the LLO CV-QKD system.

## ACKNOWLEDGMENTS

We acknowledge the financial support from the National Science Foundation of China (Grants No. 61771439, No. 61702469, No. U19A2076, and No. 61901425), the National Cryptography Development Fund (Grant No. MMJJ20170120), the Sichuan Science and Technology Program (Grants No. 2019JDJQ0060, and No. 2020YFG0289), the Sichuan Application and Basic Research Funds (Grants No. 2020YJ0482), the Innovation Special Zone Funds (18-163-00-TS-004-040-01), and the Technology Innovation and Development Foundation of China Cyber Security (JSCX2021JC001).

## APPENDIX

### A. Excess noise of the LLO pilot-multiplexed scheme

Notice that the excess noise $\xi_{tot}$ is composed of several independent and indistinguishable components. For the practical LLO pilot-multiplexed CV-QKD system, the excess noise [see Eq. (10)] could be expressed as [27]

$$\xi_{tot} = \xi_0 + \xi_{AM} + \xi_{LE} + \xi_{ADC} + \xi_{phase}. \tag{A1}$$

Therefore, the excess noise term $\xi_{rest}$ can be written as

$$\xi_{rest} = \xi_0 + \xi_{AM} + \xi_{LE} + \xi_{ADC}, \tag{A2}$$

In Eq. (A2), $\xi_0$ is the system excess noise. $\xi_{AM}$ is the modulation noise caused by Alice's modulator imperfection when preparing the Gaussian modulated coherent state signal pulse. In the LLO pilot-multiplexed scheme, since the amplitude modulator only modulates the quantum signal, the modulation noise is given by [25, 27]

$$\xi_{AM} = (E^A_{Smax})^2 10^{-\frac{d_{dB}}{10}}, \tag{A3}$$

where $E^A_{Smax} = |\alpha_{Smax}|$ is the maximal amplitude of the signal pulse to be modulated, $d_{dB}$ is the AM extinction ratio in dB. The value of $E^A_{Smax}$ can be taken as $\sqrt{10V_A}$ [25]. Due to the finite extinction ratios of the amplitude modulator and the polarization multiplexing scheme, the photon-leakage noise caused by the leakage from the phase-reference to the signal pulse can be identified as [27, 31]

$$\xi_{LE} = \frac{2(E^A_R)^2}{R_e + R_p} \tag{A4}$$

where $E^A_R = |\alpha^A_R|$ is the amplitude of the phase-reference on Alice's side, $R_e$ and $R_p$ are the finite extinction ratios of the amplitude modulator used to generate laser pulses and the polarization multiplexing scheme, respectively. As for the ADC quantization noise $\xi_{ADC}$, it is introduced by the analog-to-digital (ADC) converters imperfection. Moreover, in the LLO pilot-multiplexed scheme [27], two heterodyne detectors are used to detect the signal pulse and the phase-reference. The quantization noise related to the signal (equivalent to the excess noise at the input) only depends on the maximal amplitude of signal pulse, which satisfies [27]

$$\xi_{ADC} \geq \frac{(E^A_{Smax})^2}{12 \times 2^n} \tag{A5}$$

where n is the quantization number of the ADC.

### B. Secure key rate calculation

In the case of reverse reconciliation, the asymptotic secure key rate is given by [33]

$$K = \beta I_{AB} - \chi_{BE}. \tag{A6}$$

where $I_{AB}$ is the mutual information between Alice and Bob, $\beta$ is the reconciliation efficiency, and $\chi_{BE}$ is the Holevo information bound between Eve and Bob. For heterodyne detection, $I_{AB}$ is given by

$$I_{AB} = \log_2 \frac{V + \chi_{tot}}{1 + \chi_{tot}}. \qquad (A7)$$

with $V = V_A + 1$. The Helovo bound is expressed as

$$\chi_{BE} = \sum_{i=1}^{2} G\left(\frac{\lambda_i - 1}{2}\right) - \sum_{i=3}^{5} G\left(\frac{\lambda_i - 1}{2}\right). \qquad (A8)$$

where $G(x) = (x+1)\log_2(x+1) - \log_2 x$, and the symplectic eigenvalues stemming from the covariance matrix can be expressed as

$$\lambda_{1,2}^2 = \frac{1}{2}\left[A \pm \sqrt{A^2 - 4B}\right],$$
$$\lambda_{3,4}^2 = \frac{1}{2}\left[C \pm \sqrt{C^2 - 4D}\right],$$
$$\lambda_5 = 1. \qquad (A9)$$

where

$$A = V^2(1 - 2T) + 2T + T^2(V + \chi_{line})^2,$$
$$B = T^2(V\chi_{line} + 1)^2,$$
$$C = \frac{1}{[T(V + \chi_{tot})]^2}\left[A\chi_{het}^2 + B + 1 + 2\chi_{het} \times \left(V\sqrt{B} + T(V + \chi_{line})\right) + 2T(V^2 - 1)\right],$$
$$D = \left(\frac{V + \sqrt{B}\chi_{het}}{T(V + \chi_{tot})}\right)^2. \qquad (A10)$$

---

**Figures and Figure captions:**

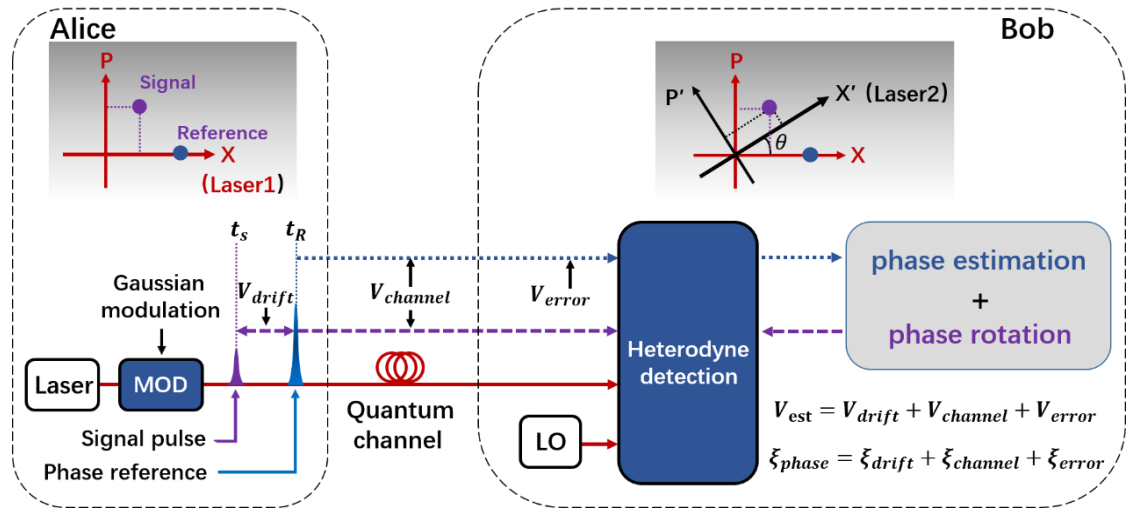

Fig. 1. Schematic for LLO CV-QKD and its phase-rotation estimation process, i.e., phase compensation. See text for details.

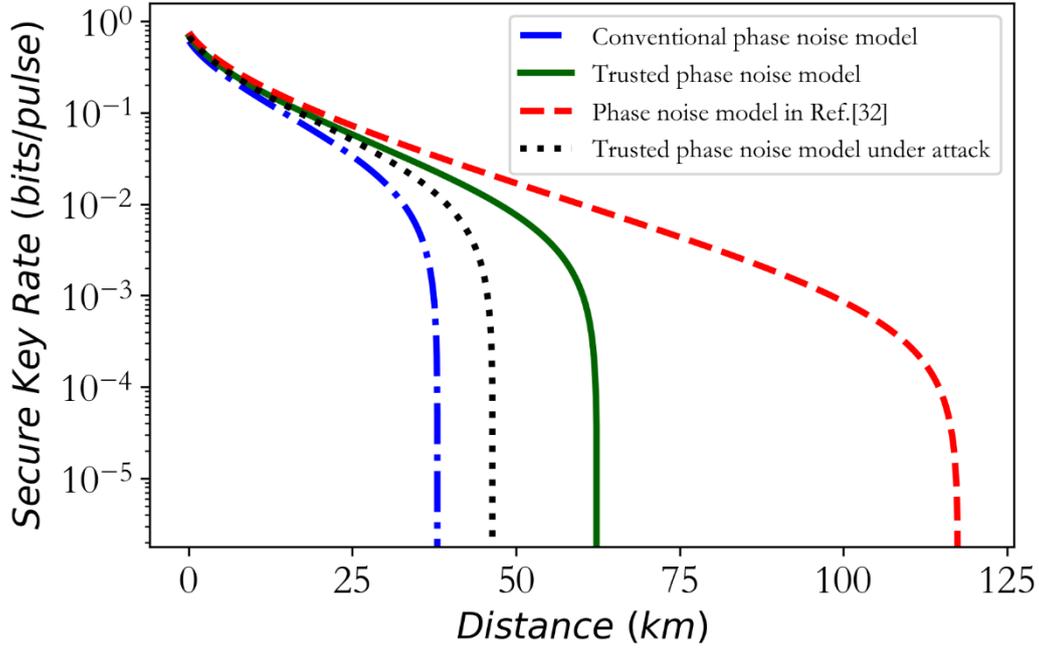

Fig. 2. (Color online). Simulation results of the secure key rate versus the transmission distance under the conventional phase noise model (blue dash-dotted line) and the trusted phase noise model (green solid line), as well as the phase noise model of Ref. [32] (red dashed line). The black dotted line shows the result of the trusted phase noise model under a practical attack scheme proposed in Ref. [32]. Some typical parameters are as follows: reconciliation efficiency $\beta = 95\%$, detector efficiency $\eta = 0.5$, modulation variance $V_A = 4$, electronic noise $v_{el} = 0.1$, attenuation coefficient $\alpha = 0.2$ dB/km, phase reference intensity $E_R^2 = 1000$, system excess noise $\xi_0 = 0.01$, ADC quantization number n=10, the AM dynamics $d_{dB} = 40$, finite extinction ratios $R_e = 40$ dB and $R_p = 30$ dB. All noise variances here are expressed in shot noise units (SNU).

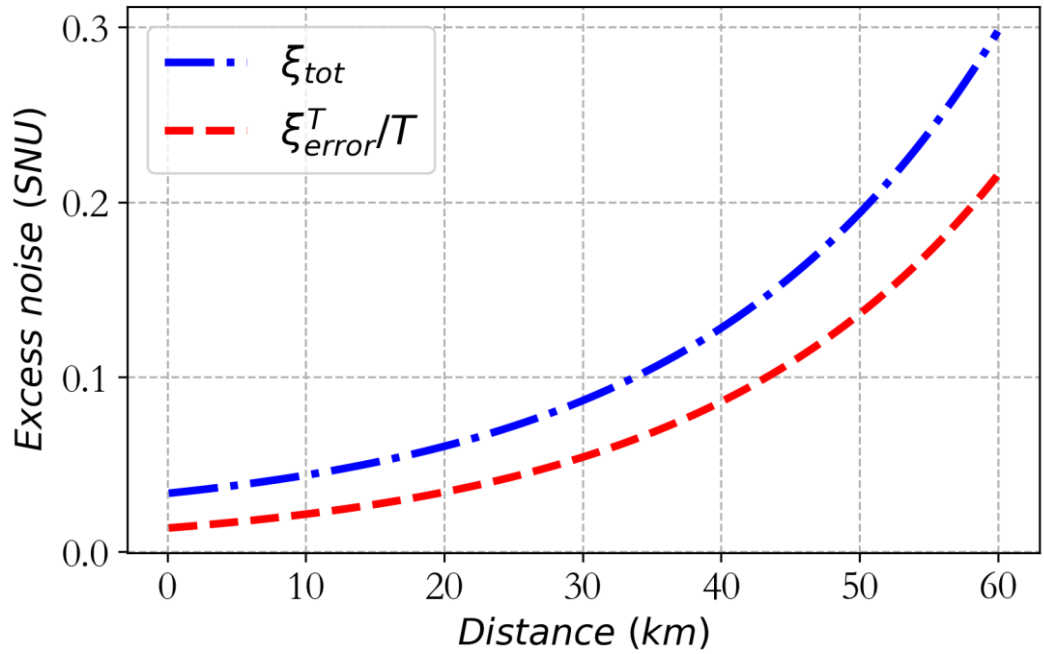

Fig. 3. (Color online). Simulation results of the excess noise and the trusted part of the phase noise referred to the channel input. The blue dash-dotted line and red dashed line respectively represent the total excess noise and the trusted part of the phase noise, referred to the channel input, versus the transmission distance. The simulated parameters are the same as that used in Fig. 2.

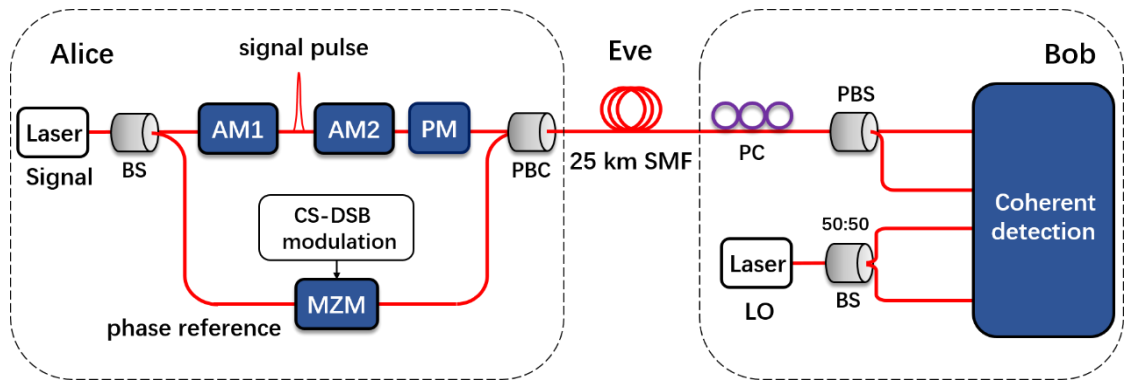

Fig. 4. A simplified schematic of the LLO pilot-tone-assisted CV-QKD system. BS, beam splitter; AM, amplitude modulator; PM, phase modulator; MZM, Mach-Zehnder modulator; CS-DSB, carrier suppression double sideband; PBC, polarization beam combiner; PC, polarization controller; PBS, polarization beam splitter; LO, local oscillator.

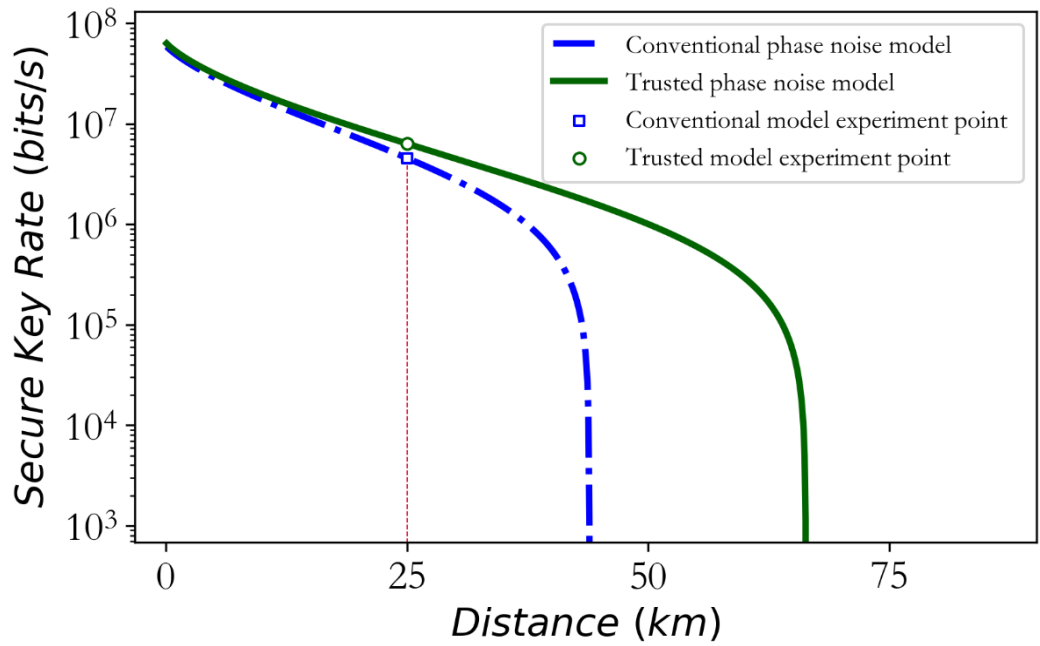

Fig. 5. (Color online). Experimental secure key rates and numerical simulations for the LLO CV-QKD system. The blue square and the green circle respectively represent the experimental results at 25 km under the conventional phase noise model and the trusted phase noise model. The blue dash-dotted line and the green-solid line respectively represent the numerical simulations of the secure key rates under the conventional phase noise model and the trusted phase noise model. The numerical simulations are calculated starting from the experimental parameters at 25 km.

TABLE I. List of the experimental parameters. $\alpha$, attenuation coefficient; $f_{\text{rep}}$, repetition frequency of the signal pulse; $\beta$, reconciliation efficiency; $\eta$, efficiency of Bob's detector; $\nu_{\text{el}}$, electronic noise; $V_A$, modulation variance; $\langle N_R \rangle$, number of average photons on Bob's side; $\xi_{\text{tot}}$, excess noise under the conventional phase noise model; $\xi_{\text{tot}}^T$, real excess noise under the trusted phase noise model; Key, secure key rate under the conventional phase noise model; $\text{Key}^T$, secure key rate under the trusted phase noise model.

| Distance(km) | $\alpha$(dB/km) | $f_{rep}$(MHz) | $\beta$ | $\eta$ | $\nu_{el}$ | $V_A$ | $\langle N_R \rangle$ | $\xi_{tot}$ | $\xi_{tot}^T$ | Key(Mbps) | $\text{Key}^T$(Mbps) |
|---|---|---|---|---|---|---|---|---|---|---|---|
| 25 | 0.2 | 100 | 0.95 | 0.56 | 0.042 | 3.073 | 1000 | 0.056 | 0.03 | 4.556 | 6.358 |